\documentclass[11pt]{article}
\usepackage{./cismylay,flushend}

\providecommand{\vctx}{{\overline x}}

\renewcommand{\eg}{{\emph{e.g.\ }}}
\providecommand{\iid}{\emph{i.i.d.}\xspace}

\newcommand{\mlt}{\mu}

\newcommand{\Atn}{{\hat A}}                

\newcommand{\pml}{{\hat p}}

\newcommand{\est}{{q}}                   

\newcommand{\sqrttpn}{{(\sqrt{2\pi})^n}}
\newcommand{\sqrttpnmo}{{(\sqrt{2\pi})^{n-1}}}
\newcommand{\sqrttp}{{\sqrt{2\pi}}}
\newcommand{\fat}{{\frac\alpha2}}

\newcommand{\pmul}{{\varphi}}
\newcommand{\frq}{\mu}

\newcommand{\Ptn}{\Psi}

\newcommand{\x}[1]{\cX({#1})}

\newcommand{\mlps}[1]{{\hat{{#1}}}}
\newcommand{\mlpptn}[1]{{\mlps{\Ptn}}}

\newcommand{\ubp}[1]{{\tilde p}}
\newcommand{\qs}{{q}}
\newcommand{\qsc}[1]{{q}}

\newcommand{\qsh}{{\qs_{1/2}}}
\newcommand{\qshc}[1]{{\qsh}}

\newcommand{\Ikns}[1]{\cI_{{#1}}^n}

\newcommand{\RwIkns}[1]{{\hat R}(\Ikns2)}

\newcommand{\estpoc}[1]{{\estpo}}
\newcommand{\esth}{{\est_{{}_{1/2}}}}
\newcommand{\esthc}[1]{\esth}

\newcommand{\pmuls}[1]{\pmul_{#1}}

\newcommand{\estpo}{{\est_{+1}}}

\newcommand{\mlts}[1]{{\mlt_{#1}}}
\newcommand{\mlti}[1]{{\mlts i}}

\newcommand{\atn}{{\Atn}}

\newcommand{\frqs}[1]{{\frq_{#1}}}
\newcommand{\frqi}[1]{{\frqs i}}

\newcommand{\pmuli}[1]{{\pmuls i}}

\providecommand{\z}{{\bf z}}
\providecommand{\ww}{{\bf w}}
\newcommand{\K}{{\bf K}}
\newcommand{\C}{{\bf C}}
\renewcommand{\x}{{\bf x}}
\renewcommand{\u}{{\bf u}}
\newcommand{\varmm}[3]{\Paren{#1}_{#2}^{#3}}

\title{Universal compression of Gaussian sources with unknown parameters}

\author{Alon Orlitsky${}^{\dagger}$ and Narayana Santhanam${}^*$
}

 \ignore{Such \emph{universal
      descriptions} for the random variable are agnostic to the identity of the
    distribution in $\cP$.}

\begin{document}
\maketitle
\footnote{
${}^{\dagger}$Department of Electrical and Computer Engineering, University of California, San Diego \\
Email: \texttt{alon@ucsd.edu}\\
${}^*$Department of Electrical Engineering, University of Hawaii at Manoa\\ Email: \texttt{nsanthan@hawaii.edu}}
\begin{abstract} For a collection of distributions over a countable
  support set, the worst case universal compression formulation by
  Shtarkov attempts to assign a universal distribution over the
  support set. The formulation aims to ensure that the universal
  distribution does not underestimate the probability of any element
  in the support set relative to distributions in the collection. When
  the alphabet is uncountable and we have a collection $\cP$ of
  Lebesgue continuous measures instead, we ask if there is a
  corresponding universal probability density function (pdf) that does
  not underestimate the value of the density function at any point in
  the support relative to pdfs in $\cP$.  An example of such a measure
  class is the set of all Gaussian distributions whose mean and
  variance are in a specified range. We quantify the formulation in
  the uncountable support case with the \emph{attenuation} of the
  class---a quantity analogous to the worst case redundancy of a
  collection of distributions over a countable alphabet. An
  attenuation of $A$ implies that the worst case optimal universal pdf
  at any point $x$ in the support is always at least the value any pdf
  in the collection $\cP$ assigns to $x$ divided by $A$.  We analyze
  the attenuation of the worst optimal universal pdf over length-$n$
  samples generated \iid from a Gaussian distribution whose mean can
  be anywhere between $-\alpha/2$ to $\alpha/2$ and variance between
  $\sigma_m^2$ and $\sigma_M^2$. We show that this attenuation is
  finite, grows with the number of samples as $\cO(n)$, and also
  specify the attentuation exactly without approximations. When only
  one parameter is allowed to vary, we show that the attenuation grows
  as $\cO(\sqrt{n})$, again keeping in line with results from prior
  literature that fix the order of magnitude as a factor of $\sqrt{n}$
  per parameter. In addition, we also specify the attenuation exactly
  without approximation when only the mean or only the variance is
  allowed to vary.  
\end{abstract}

Compression has been well studied since Shannon~\cite{sha48} formalized not just
the notion of what it means to represent data or signals in a compact form, but
also quantified how compact the representation can be. For data that come from a
countable (discrete) alphabet, this lower bound on compression is essentially
the entropy of the source. Furthermore, concrete schemes to represent discrete
data in bits are also known---the Huffman coding scheme being the optimal one.

While the quantification of the limits of compression is elegant, it
does not take into account one of the practicalities of
compression---we do not know the underlying distribution. 
\ignore{Hence we
do not know the underlying entropy. Since the underlying distribution
is unknown, it has to be infered through the data.  Here we run into
trouble. The limit of compression, entropy, is not a well behaved
function of the underlying distribution if the underlying support is
not a finite set. It is easy to see
that distributions arbitrarily close to each other in, say the
$\ell_1$ metric, could have their entropies arbitrarily far from
each other--see,~\eg~\cite{SA12:spcom}.}
Instead, Shtarkov~\cite{sht87} considered a compression framework
where the underlying distribution remains unknown, assuming instead
that the unknown distribution belongs to a known collection $\cP$ of
possible distributions.  The framework in~\cite{sht87} is a natural
approach to \emph{universal compression}~\cite{fit72}, where we attempt
to describe the data almost as well as the underlying model by means of a
\emph{universal} distribution. Suppose we have a class $\cP$ of
distributions over a countable set $X$.  We now attempt to
find a universal distribution $q$ over $X$ such that
\begin{equation}
\label{eq:rdn}
\sup_{x\in X} 
\sup_{p\in\cP}
\frac{p(x)}{q(x)}
\end{equation}
is as small as possible. The ratio above is always $\ge 1$, since for any two
distributions $p$ and $q$ over $X$
\[
\sup_{x\in X} 
\frac{p(x)}{q(x)} \ge 1.
\]
Suppose the supremum in~\eqref{eq:rdn} is finite and
equals $A$. It follows that no matter what the realization $x$ is
or the underlying model $p$ is
\[
q(x)\ge \frac{p(x)}A.
\]
Therefore, where $A$ is suitably close to 1, the universal $q$ obviates the need
to know the underlying distribution $p$ within $\cP$.

When we deal with sequences of variables, the quantity of interest is
often not the entire sequence itself. Rather, we may be interested in
different statistics of a sequence. For example, in the \iid case, the
sum of the sequence of random variables is a sufficient statistic.

There are several large deviation results that help us tackle such
statistics better. Indeed, large deviation results for sums of many
kinds of sequences of random variables are well known. At the very
simplest, the sum of \iid Bernoulli random variables
is distributed as Gaussian in the 
limit as the number of variables increases to infinity.
The mean of the Gaussian
equals to the mean of the Bernoulli random variable and its variance
is easily related to the Bernoulli variable's variance.

More generally, the limit probability law need not always be Gaussian
as above even when we consider the component random variables to be
binary. With appropriate scaling of probabilities of the individual
binary random variables, we can have the limiting law be Poisson for
example~\cite[vol 2, p173]{Fel66}. Indeed, the different distributions
that could come up as the limiting law of sums of random variables are
characterized as \emph{infinitely divisible} distributions
(see~\eg~\cite[vol 2, ch 6]{Fel66} or~\cite{sat99}). This family of
infinitely divisible distributions includes several well known
distributions such as the negative binomial, Gamma, $\chi^2$ and
Cauchy distributions, in addition to Gaussians and Poisson
distributions. In all these cases, it is natural to use the limiting
infinitely divisible distribution to describe the sum, rather than the
sequence of random variables.

For more general functions other than the sum, deviation bounds such
as Hoeffding's inequality or McDiarmid's inequality (among others)
allow us to define a dominating distribution on the deviation of the
function from its mean value. Often, the mean of these dominating
distributions is easily obtained and the general fall off of large
deviations. Describing these functions is therefore better handled by
describing the dominating distribution rather than the sequence
itself.

In both cases---whether we consider infinitely divisible distributions
or distributions that characterize large deviations, we may have to
deal with a family of distributions with uncountable support such as
Gaussian distributions. The exact parameters of the distribution in
question is a function of the underlying statistics of the sequence
though the family the distribution belongs to is fixed to within the
range of parameters. The natural question then is, in analogy with how
we deal with countable data, can we universally handle these collections
of distributions over uncountable supports as well?

Let $X$ be an uncountable set, say the real line.  Suppose, as before,
that $\cP$ is the collection of probability measures over $X$. 
In addition, the measures in $\cP$ are absolutely continuous with
respect to the Lebesgue measure. We see data from an unknown measure
in $\cP$. Could we take a universal approach again and come up with a
universal pdf for $\cP$ that does not underestimate the true density
anywhere?

Surprisingly, despite the strong motivation, the uncountable support
case has received very little attention, despite the seminal work of
Rissanen~\cite{Ris86}.  The multitude of
results~\cite{kt81,xb00,ky00,ds02} on universal compression over
finite alphabets do not apply non-trivially when the domain is
uncountable.

One exception is Rissanen's results in~\cite{Ris86} that indicates
that even while the support may be uncountable, if the class $\cP$ of
probability measures can be parameterized by a few parameters there
must be a good universal measure for $\cP$. Formally, we define the
attenuation of a collection $\cP$ of measures over the a support
$X\subseteq \reals$ in analogy with Shtarkov~\cite{sht87} and
Rissanen~\cite{Ris86}. Suppose every measure in the collection $\cP$
is absolutely continuous with the Lebesgue measure for the sake of
simplicity. Then, we define the attenuation
\[
\atn(\cP)=
\inf_{q}
\sup_{x\in X} 
\sup_{p\in\cP}
\frac{p(x)}{q(x)},
\]
where $p\in\cP$ and $q$ are the probability distribution functions
(pdfs) with respect to the Lebesgue measure defined in the standard
way. We also let for all $x\in X$,
\[
\hat{p}(x)=\sup_{p\in\cP} p(x).
\]
\bRemark The requirement of absolute continuity with the Lebesgue
measure can be relaxed in several ways.  One way is to decompose
measures into a discrete probability distribution and a probability
density function. It is also possible to have a more general (and
cleaner, if more abstract) formulation where we simply require all
$p\in\cP$ to be absolutely continuous with respect to the universal
$q$, and consider the Radon-Nikodym derivative in place of pdf
ratios. However, we keep the restriction in this paper to focus on
Gaussian probability density functions.  \eRemark

To make the problem concrete, we consider collections of Gaussian
distributions over the real line. If, as in the Gaussian case,
$\hat{p}(x)$ is measurable we clearly have
\[
\atn(\cP) =\int_{X} \hat{p}(x) dx.
\]
If the integral above is bounded, we say that the \emph{attenuation}
of $\cP$ is finite. Here the pdf $q^*$ that achieves the infimum in
the definition of attenuation above is easily seen to be
\[
q^*(x)=\frac{ \hat{p}(x)}{\int_{x'\in X} \hat{p}(x') dx'}.
\]
In particular, we also consider the case where $X$ is the space of
sequences of real numbers (sampled \iid) from distributions in $\cP$.

This paper studies collections of Gaussian distributions over real
numbers.  As mentioned before, Gaussian distributions form the limit
law of sums of a wide variety of \iid random
variables---see~\cite{Fel66,Kal97} for more details. When the
individual random variables can be from alphabets other than binary,
it is possible to characterize the mean of the limit law without
knowing the variance, and vice versa. Furthermore, we also study the
attenuation of sequences of \iid Gaussian random
variables---corresponding to describing disjoint partial sums of a
sequence of an unknown \iid random variables.

We consider two cases. In the first case, only one parameter
(either the mean or the variance) is unknown while the other is
specified. These results appear in Theorem~\ref{thm:galpha} and
Corollary~\ref{corr:variance}. The second case allows both the mean
and variance to be unknown, and is treated in Theorems~\ref{thm:app}
and~\ref{thm:exact}. In both cases, we will also calculate the
attenuation of \iid sampling from the Gaussian collection precisely,
without any approximations. These results extend and make more precise
a specific section on results on similar collections in~\cite{Ris86}.

A word on notation---we will use bold font to denote vectors and
matrices.  The transpose of a matrix $\K$ is $\K^T$ and its determinant is $|\K|$. For a vector
$\x=(x_1\upto x_n)$, we use $d\x$ to denote $dx_1\ldots dx_n$. We will
interchangeably refer to length-$n$ sequences $x_1\upto x_n$ by 
their length-$n$ column vector analogs $\x=(x_1\upto x_n)^T$.

\section{Gaussians with unknown mean and variance 1}
Let $G_\alpha$ be the collection of gaussians with variance
$\sigma^2=1$ and where the mean $\mu$ lies in the range
$-\alpha/2\le\mu\le\alpha/2$ (total range is $\alpha$).  We denote by
$G_{\alpha}^n$ the collection of all pdfs on $\reals^n$ obtained by
\iid sampling from a distribution in $G_{\alpha}$.  We will do a
couple of examples before computing the attenuation for length-$n$
strings from the class $G^n_\alpha$ for general $n$. In this section,
for any length-$n$ sequence $\x$, we denote $\hat p(\x)=\arg\max_{p\in
  G_\alpha^n}p(\x)$.

\bExample
\label{eg:galphaone} (Length 1) If $-\alpha/2\le x\le\alpha/2$, the Gaussian in
$G_\alpha$ maximizing $p(x)$ has mean $x$, hence $\hat
p(x)=\frac1{\sqrt{2\pi}}$.  If $x\ge\alpha/2$, the Gaussian in
$G_\alpha$ maximizing $p(x)$ has mean $\alpha/2$, hence $\hat
p(x)=\frac1{\sqrt{2\pi}}e^{-\half(x-\alpha/2)^2}$.  Similarly for
$x\le -\alpha/2$.

The attenuation for a sequence of length 1 is therefore
\begin{align*}
\atn(G^1_\alpha)
&=\int_{-\infty}^\infty \hat p(x)dx\\
&=
\int_{-\infty}^{-\alpha/2} \hat p(x)dx+
\int_{-\alpha/2}^{\alpha/2} \hat p(x)dx+
\int_{\alpha/2}^\infty \hat p(x)dx\\
&=\half+\frac\alpha{\sqrt{2\pi}}+\half\\
&=1+\frac\alpha{\sqrt{2\pi}},
\end{align*}
which makes sense as if $\alpha=0$, we know the distribution and have
attenuation 1. 
\eExample

Next consider attenuation for sequences of length 2.

\bExample 
Let $x_1$ and $x_2$ denote the first and second outcomes. 
Define $y=(x_1+x_2)/2$ to be the average and $z=x_1-y=(x_1-x_2)/2$ to be 
the difference between $x_1$ and the average. 
A gaussian with mean $\mu$ will assign the sequence $(x_1,x_2)$ probability
\[
p(x_1,x_2)=
\frac1{2\pi}e^{-\half[(x_1-\mu)^2+(x_2-\mu)^2]}
\]
Therefore the maximum likelihood (ML) Gaussian in $G^2_\alpha$, $\hat
p$, has mean $\mu=y$ if $-\alpha/2\le y\le\alpha/2$, has
$\mu=\alpha/2$ if $y>\alpha/2$, and $\mu=-\alpha/2$ if $y<-\alpha/2$.

It follows that the attenuation for 2-element sequences is
\begin{align*}
\atn(G^2_\alpha)
&=\int_{-\infty}^\infty dx_1\ \int_{-\infty}^\infty dx_2\ \hat p(x_1,x_2)\\
&=2\int_{-\infty}^\infty dy\ \int_{-\infty}^\infty dz\ \hat p(y,z)\\
&=\frac2{(\sqrt{2\pi})^2}
\int_{-\alpha/2}^{\alpha/2}dy\ \int_{-\infty}^\infty dz\ \exp\Paren{-\half z^2-\half z^2}\\
&\quad+
\frac{2\cdot 2}{(\sqrt{2\pi})^2}\int_{\alpha}^\infty dy\int_{-\infty}^\infty dz\
\exp-\half\Paren{\Paren{\frac\alpha2-(y-z)}^2+\Paren{\frac\alpha2-(y+z)}^2}.
\end{align*}
Now, the first summand is 
\[
\frac{\sqrt2}{\sqrt{2\pi}}
\int_{-\alpha/2}^{\alpha/2}dy\ 
\frac{\sqrt2}{\sqrt{2\pi}}\int_{-\infty}^\infty dz \exp\Paren{-z^2}
=
\frac{\sqrt2}{\sqrt{2\pi}}\int_{-\alpha/2}^{\alpha/2}dy
=\frac\alpha{\sqrt{\pi}},
\]
and the second summand is 
\[
\frac{2\sqrt2}{\sqrt{2\pi}}
\cdot \int_{\alpha/2}^\infty dy
\frac{\sqrt2}{\sqrt{2\pi}}
\int_{-\infty}^\infty dz\ 
\exp-\Paren{\Paren{y-\frac{\alpha}2}^2+z^2}
=
1.
\]
So
\[
\atn(G^2_\alpha)=1+\frac\alpha{\sqrt{\pi}}.\eqed
\]
\eExamplep

\bTheorem
\label{thm:galpha}
For all $n$,
\[
\atn(G^n_\alpha)=1+\alpha\sqrt{\frac n{2\pi}}.
\]
\Proof The case $n=1$ has been considered in
Example~\ref{eg:galphaone}. For $n\ge2$, we will transform the length
$n$ sequence $\x=(x_1\upto x_n)^T$ into the following variables
\[
y=(x_1+\ldots+x_n)/n, \text{ and }
z_j=x_j-y \text{ for }1\le j\le n-1.
\]
Now $\z=(z_1\upto z_{n-1})^T$ takes values in $\reals^{n-1}$ and $y\in\reals$.
Then the Jacobian of the transformation,
\[
\frac{\partial y\z}{\partial \x}
=
\frac1n
\left(
\begin{array}{rrrcrr}
1   & 1   & 1  & \ldots & 1 & 1\\
n-1 & -1  & -1 & \ldots & -1 & -1\\
-1  & n-1 & -1 & \ldots & -1 & -1\\
-1  & -1 & n-1 &  & -1 & -1\\
\vdots & \vdots & & \ddots & \vdots &\vdots\\
-1  & -1  & -1 & \ldots & n-1 & -1
\end{array}
\right),
\]
and its determinant,
\begin{equation}
\label{eq:jac}
\left|\frac{\partial y\z}{\partial \x}\right| 
= (-1)^{n-1}\frac{\Paren{(n-1)+1}^{n-1}}{n^n}
=\frac{(-1)^{n-1}}n.
\end{equation}
We will compute the attenuation using the above transformation. The length-$n$ attenuation,
\begin{align*}
\atn(G^n_\alpha)
&=
\int\pml(\vctx)d\x\\
&= n \int\pml(y,z_1\upto z_{n-1})dyd\z\\
&= 
2\frac{n}{\sqrttpn}
\int_{y=\alpha/2}^\infty\int_{z_1\upto z_{n-1}=-\infty}^\infty
\!\!\!\!\!\!\!\!\!\!\!\!
\exp\Paren{ -\frac{
\sum_{i=1}^{n-1}\Paren{\fat-(y+z_i)}^2+\Paren{\fat-(y-\sum_{i=1}^{n-1}z_i)}^2}2}
dyd\z\\
&\qquad\qquad\qquad+
\frac{n}{\sqrttpn}
\int_{y=-\alpha/2}^{\alpha/2}\int_{z_1\upto z_{n-1}=-\infty}^\infty
\!\!\!\!\!\!\!\!\!\!\!\!
\exp\Paren{ -\frac{
\sum_{i=1}^{n-1}z_i^2+\Paren{\sum_{i=1}^{n-1}z_i}^2}2}
dyd\z.
\end{align*}
We simplify the first integral in the last line above using
\[
\sum_{i=1}^{n-1}\Paren{\fat-(y+z_i)}^2+\Paren{\fat-(y-\sum_{i=1}^{n-1}z_i)}^2\\
=
n\Paren{\fat-y}^2+\sum_{i=1}^{n-1}z_i^2+\Paren{\sum_{i=1}^{n-1}z_i}^2.
\]
Doing so, and letting
\[
I
\ed
\frac{\sqrt n}{\sqrttpnmo}
\int_{z_1\upto z_{n-1}=-\infty}^\infty
\!\!\!\!\!\!\!\!\!\!\!\!
\exp\Paren{ -\frac{
\sum_{i=1}^{n-1}z_i^2+\Paren{\sum_{i=1}^{n-1}z_i}^2}2} d\z,
\]
we obtain
\[
\atn(G^n_\alpha)
=
I\cdot\Paren{
2\frac{\sqrt n}{\sqrttp}
\int_{y=\alpha/2}^\infty
\exp -\half\Paren{n\Paren{\fat-y}^2}dy
+\frac{\sqrt n\alpha}{\sqrttp}}
=
I\cdot\Paren{1+\alpha\sqrt{\frac n{2\pi}}}.
\]
We will now show that the integral $I=1$ to conclude the proof of the
theorem. Write
\begin{equation}
\label{eq:zkinv}
\sum_{i=1}^{n-1}z_i^2+\Paren{\sum_{i=1}^{n-1}z_i}^2
=
\sum_{i=1}^{n-1}2z_i^2+\sum_{1\le i,j\le n-1}z_iz_j
=
\z^T \K^{-1}\z
\end{equation}
where $\K^{-1}$ is a $(n-1)\times (n-1)$ matrix, given by
\[
\K^{-1}
=
\left(
\begin{matrix}
2 & 1 & 1 & \cdots & 1\\
1 & 2 & 1 & \cdots & 1\\
1 & 1 & 2 &        & 1\\
\vdots & \vdots & & \ddots & \vdots\\
1 & 1 & 1 & \cdots & 2\\
\end{matrix}
\right).
\]
Now letting ${\mathbf I}_{n-1}$ be an identity matrix of dimensions $(n-1)\times(n-1)$ and
writing $\mathbf 1$ for a column vector of $n-1$ ones, we have
\[
\K^{-1}=
{\mathbf I}_{n-1}+ {\mathbf 1}{\mathbf 1}^T.
\]
It follows from the Syvelster determinant theorem~\cite{syl:tt} that
\begin{equation}
\label{eq:kinv}
|\K^{-1}|=|{\mathbf I}_{n-1}+ {\mathbf 1}{\mathbf 1}^T| 
= 
|1+ 
{\mathbf 1}^T{\mathbf 1}|
=
1+(n-1)=n.
\end{equation}
Hence
\[
|\K|=\frac1n.
\]
Therefore, 
\begin{align}
\nonumber
I
&=
\frac{\sqrt n}{\sqrttpnmo}
\int_{z_1\upto z_{n-1}=-\infty}^\infty
\!\!\!\!\!\!\!\!\!\!\!\!
\exp\Paren{ -\frac{
\sum_{i=1}^{n-1}z_i^2+\Paren{\sum_{i=1}^{n-1}z_i}^2}2}d\z\\
\nonumber
&=
\frac{1}{\sqrttpnmo\sqrt{|\K|}}
\int_{z_1\upto z_{n-1}=-\infty}^\infty
\exp\Paren{ -\frac{\vctz^T \K^{-1}\vctz}2}d\z\\
\label{eq:inti}
&=1.
\end{align}
The theorem follows.
\eTheorem

\section{Gaussians with unknown mean and variance}
Let $G_{\alpha,\sigma_m,\sigma_M}$ be the collection of iid gaussians
with $-\alpha/2\le\mu\le\alpha/2$ and $\sigma_m\le\sigma\le\sigma_M$.
Throughout this section, we will use
\[
p_{\sigma,\mu}(x)=\frac1{\sqrt{2\pi}\sigma}\exp\Paren{-\frac{(x-\mu)^2}{2\sigma^2}}
\]
to denote a Gaussian with variance $\sigma^2$ and mean $\mu$. As before,
we denote the collection of all pdfs on $\reals^n$ obtained by \iid sampling from a 
distribution in $G_{\alpha,\sigma_m\sigma_M}$ by $G_{\alpha,\sigma_m\sigma_M}^n$.
As before, in this section, for any length-$n$ sequence $\x$, we  
denote $\hat p(\x)=\arg\max_{p\in G_{\alpha,\sigma_m,\sigma_M}^n}p(\x)$.

\bExample
\label{eg:mv}
Let 
\[
p_\sigma(x)=\frac1{\sqrt{2\pi}\sigma}\exp\Paren{-\frac{x^2}{2\sigma^2}}
\]
denote the Gaussian distribution with zero mean and standard deviation 
$\sigma$. 
Differentiating $\log p_\sigma(x)$ with respect to $\sigma$, we obtain 
that for every $x$, $p_\sigma(x)$ is maximized by $\sigma=x$. 
Therefore
\[
\max_\sigma p_\sigma(x)=\frac1{\sqrt{2\pi}x}\exp\Paren{-\half}
=
\frac1{\sqrt{2\pi e}x}.
\]
It follows that 
\[
\pml(x)
=
\begin{cases}
\frac1{\sqrt{2\pi}\cdot\sigma_m} &0\le|x|\le\frac\alpha2\\
\frac1{\sqrt{2\pi}\cdot\sigma_m}\exp\Paren{-\half\frac{\paren{|x|-\alpha/2}^2}{\sigma_m^2}}
& \frac\alpha2\le|x|\le\frac\alpha2+\sigma_m\\
\frac1{\sqrt{2\pi e}\cdot(|x|-\alpha/2)} & 
\frac\alpha2+\sigma_m\le |x|\le \frac\alpha2+\sigma_M\\
\frac1{\sqrt{2\pi}\cdot\sigma_M}\exp\Paren{-\half\frac{\paren{|x|-\alpha/2}^2}{\sigma_M^2}}
& \frac\alpha2+\sigma_M\le|x|.\\
\end{cases}
\]
Hence
\[
\atn(G^1_{\alpha,\sigma_m,\sigma_M})
=
1+
\frac\alpha{\sigma_m}\cdot\sqrt{\frac1{2\pi}}
+
\sqrt{\frac2{\pi e}}\cdot\ln\frac{\sigma_M}{\sigma_m}.\eqed
\]
\eExamplep
Given a sequence $\x=(x_1\upto x_n)^T$, we let as before
\[
y=\frac{\sum_{i=1}^n x_i }n. 
\]
Furthermore for $a_-$, $a_+$ and $A$ in $\reals$, let
\[
\varmm{A}{a_-}{a_+}
=
\begin{cases}
a_- & A\le a_-\\
A & a_-\le A\le a_+\\
a_+ & a_+\le A.
\end{cases}
\]
The following lemma characterizes the maximum likelihood distribution.
\bLemma
\label{lm:mlesm}
For $x_1\upto x_n\in \reals^n$, the ML estimates of the mean and variance are
\[
\hat\mu= \varmm{y}{-\alpha/2}{\alpha/2}
\]
and 
\[
\hat\sigma^2=\varmm{\frac{\sum_{i=1}^n (x_i-\hat\mu)^2}n}{\sigma_m^2}{\sigma_M^2}.
\]
Namely, 
\[
\arg\max_{\sigma,\mu} p_{\sigma,\mu}(x_1\upto x_n) = p_{\hat\sigma,\hat\mu}.
\]
\Proof
\newcommand{\psm}{p_{\sigma,\mu}}
If 
\[
\varmm{y}{-\alpha/2}{\alpha/2}=y \text{ and }
\varmm{\frac{\sum_{i=1}^n (x_i-\hat\mu)^2}n}{\sigma_m^2}{\sigma_M^2}
= \frac{\sum_{i=1}^n (x_i-\hat\mu)^2}n
\]
the lemma follows by noting that the first partial derivatives of $\psm$ are zero at $\hat\mu=y$ 
and $\hat\sigma=\frac{\sum_{i=1}^n (x_i-y)^2}n$. The second partial derivatives 
can be easily verified to satisfy
\[
\frac{\partial^2 \psm}{\partial \sigma\partial \mu}
-
\frac{\partial^2 \psm}{\partial\sigma^2}
\frac{\partial^2 \psm}{\partial\mu^2}
< 0
\]
with $\frac{\partial^2 \psm}{\partial\sigma^2} <0$ and
$\frac{\partial^2 \psm}{\partial\mu^2}< 0$, meeting the conditions for
a maxima of $\psm$. If $\varmm{y}{-\alpha/2}{\alpha/2}\ne y $, the
corresponding first derivative of $\psm$ at $\mu=\hat\mu$ is
non-zero. Therefore, moving to the interior of the parameter space
parallel to the direction of a unit vector along $\mu$ decreases
$\psm$. Hence $\psm$ must be maximized on the boundary. A similar
observation holds for $\sigma$ as well. The lemma follows.  \eLemmap

Finally, we will need Stirling's approximation of the Gamma function.
\bLemma
\label{lm:str}
(Stirling) $\Gamma(x+1) = \sqrt{2\pi x} \Paren{\frac{x}{e}}^x \Paren{1+\cO\Paren{\frac1x}}$.
\eLemma

We are now in a position to compute the attenuation of $G^N_{\alpha,\sigma_m,\sigma_M}$. The 
main theorem below, Theorem~\ref{thm:app} uses the Stirling's approximation to simplify
results into a easily readable form. Theorem~\ref{thm:exact} gives the same result in
precise form, though it is unwieldy. 

\bTheorem
\label{thm:app}
For $n\ge2$,
\[
\atn(G_{\alpha,\sigma_m,\sigma_M}^n)=
\frac{\alpha\sqrt{n(n-1)}}{\pi\sqrt{2}}\Paren{\frac1{\sigma_m}-\frac1{\sigma_M}}
+\frac{\alpha\sqrt{n}}{\sqrt{2\pi}}
\Paren{\frac{I_n }{\sigma_m}+\frac{1-I_n}{\sigma_M}}
+\sqrt{\frac{n}\pi}\ln\frac{\sigma_M}{\sigma_m}+\cO(1),
\]
where
\[
I_n
\ed
\frac{\sqrt n}{\sqrttpnmo}
\int_{\substack{\sum_{i=1}^{n-1} z_i^2 +(\sum_{j=1}^{n-1} z_j)^2\le n}}
\!\!\!\!\!\!\!\!\!\!\!\!
\exp\Paren{ - \frac{\sum_{i=1}^{n-1}z_i^2+\Paren{\sum_{i=1}^{n-1}z_i}^2}2}d\z.
\]
As $n\to\infty$, we have $I_n\to 1$.
\Proof Denote $\x=(x_1\upto x_n)^T$. We compute the integral
\[
\int_{\x} \hat{p}(\x) d\x
\]
by splitting the domain of the integral, first based on the value of the mean followed by the 
value of the ML estimate of the variance. Specifically, we partition 
\[
\reals^n = \cR_1\union\cR_2\union \cR_3,
\]
where $\cR_1\ed \sets{ \x : \x^T\mathbf1 \le -n\alpha/2}$,
$\cR_2\ed \sets{ \x : -n\alpha/2\le\x^T\mathbf1 \le n\alpha/2}$,
and $\cR_3\ed \sets{ \x : \x^T\mathbf1 \ge n\alpha/2}$.  We
consider each of the regions separately below. Regions $\cR_1$ and
$\cR_3$ contribute $\cO(\sqrt{n})$ terms each, while $\cR_2$
contributes $\cO(n)$. This is to be expected since both parameters are
in play in $\cR_2$ while only one (the variance) is
effectively in play in $\cR_1$ and $\cR_3$.

\paragraph*{Region $\cR_1$}
In this region, the ML estimate of the mean is $-\alpha/2$. Depending on the ML
estimate of the variance, we further subdivide
\[
\cR_1=\cR_{11}\union \cR_{12} \union \cR_{13},
\]
where 
\begin{align*}
\cR_{11}&\ed \sets{ \x \in\cR_1: (\x+\alpha/2)^T (\x+\alpha/2)\le n\sigma_m^2}, \\
\cR_{12}&\ed \sets{ \x \in\cR_1: n\sigma_m^2\le (\x+\alpha/2)^T (\x+\alpha/2)\le n\sigma_M^2}, \\
\cR_{13}&\ed \sets{ \x \in\cR_1: (\x+\alpha/2)^T (\x+\alpha/2)\ge n\sigma_M^2}.
\end{align*}
From Lemma~\ref{lm:mlesm}, we have
\begin{align*}
\int_{\cR_{11}} \hat{p}(\x)d\x&+\int_{\cR_{13}} \hat{p}(\x)d\x\\
&=\frac1{(2\pi)^{n/2}\sigma_m^2}
\int_{\x\in\cR_{11}}
\exp\Paren{-\frac{\sum_{i=1}^n (x_i+\alpha/2)^2}{2\sigma_m^2}}d\x\\
&\quad+
\frac1{(2\pi)^{n/2}\sigma_M^2}
\int_{\x\in\cR_{13}}
\exp\Paren{-\frac{\sum_{i=1}^n (x_i+\alpha/2)^2}{2\sigma_M^2}}d\x\\
&\aeq{(a)}
\frac1{(2\pi)^{n/2}}
\int_{\substack{\u\in\reals^n\\\u^T{\mathbf1}\ge 0}}
\exp\Paren{-\frac{\u^T\u}2}d\u\\
&=
\frac12.
\end{align*}
In the above, we obtain $(a)$ by transforming the variables in the first integral using $u_i=(x_i+\alpha/2)/\sigma_m$ 
and the second integral using $u_i=(x_i+\alpha/2)/\sigma_M$. Note as before that $\u=(u_1\upto u_n)^T$.
Meanwhile, 
\begin{align}
\nonumber
\int_{\cR_{12}} \hat{p}(\x)d\x &\aeq{(a)}
\int_{
\substack{\u^T{\mathbf1}\ge 0\\
\sigma_m^2\le \u^T\u/n\le \sigma^2_M}}
\frac{n^{n/2} e^{-n/2} d\u}{(2\pi)^{n/2}(\u^T\u)^{n/2} }\\
\label{eq:one}
&\aeq{(b)}
\frac{n^{n/2} e^{-n/2}}{(2\pi)^{n/2}}
\frac{n\pi^{n/2}}{2\Gamma\Paren{\frac{n}2+1}}
\int_{r=\sqrt{n}\sigma_m}^{\sqrt{n}\sigma_M}\frac1r dr
\\
\nonumber
&=\half\sqrt{\frac{n}{\pi}}
\ln\frac{\sigma_M}{\sigma_m} +\cO\Paren{\frac1{\sqrt{n}}}\\
\nonumber
&\sim
\half\sqrt{\frac{n}{\pi}}
\ln\frac{\sigma_M}{\sigma_m}.
\end{align}
In the above, we get $(a)$ by transforming $u_i= x_i+\alpha/2$. To see
$(b)$, we transform $\u$ into polar coordinates and note (\eg~\cite{bal97})
that the surface area of a $n-$dimensional unit sphere is
\[
\frac{n\pi^{n/2}}{\Gamma\Paren{\frac{n}2+1}},
\]
while the surface area corresponding to $\u^T{\mathbf1}\ge0$ is exactly half the
above quantity. The next equality
follows because Stirling's approximation for the Gamma function above has a multiplicative accuracy of $(1+\cO\Paren{\frac1n})$
as in Lemma~\ref{lm:str}.
The net contribution to the attenuation from region $\cR_1$ is therefore
\[
\half + \half\sqrt{\frac{n}{\pi}}
\ln\frac{\sigma_M}{\sigma_m}+\cO\Paren{\frac1{\sqrt{n}}}.
\]
\paragraph*{Region $\cR_3$} This region contributes an identical amount as $\cR_1$ above. \ignore{Therefore the contribution 
from $\cR_3$ is
\[
\half + \half\sqrt{\frac{n}{\pi}}
\ln\frac{\sigma_M}{\sigma_m}+\cO\Paren{\frac1{\sqrt{n}}}.
\]}
\paragraph*{Region $\cR_2$} 
To tackle this region, we will need the auxillary variable
\[
y= \frac{\sum_{i=1}^n x_i}n, 
\]
while we will also define auxillary variables very similar to $z_j$ from Theorem~\ref{thm:galpha}.
Once again, we partition 
\[
\cR_2=\cR_{21}\union \cR_{22}\union \cR_{23},
\]
with
\begin{align*}
\cR_{21} &= \sets{ \x\in\cR_2 : \sum_{i=1}^n (x_i-y)^2 \le n \sigma_m^2},\\
\cR_{22} &= \sets{ \x\in\cR_2 : n\sigma_m^2\le \sum_{i=1}^n (x_i-y)^2 \le n \sigma_M^2},\\
\cR_{23} &= \sets{ \x\in\cR_2 : n\sigma_M^2\le \sum_{i=1}^n (x_i-y)^2 }.
\end{align*}
We will first consider the regions $\cR_{21}$ and $\cR_{23}$. We will
focus on the case $n\ge2$ here since the case $n=1$ has already been
handled by Example~\ref{eg:mv}. The contribution to the attenuation
of regions $\cR_{21}$ and $\cR_{23}$ is
\[
\int_{\x\in\cR_{21}}
\frac1{(2\pi)^{n/2} \sigma_m^n}
\exp\Paren{-\frac{\sum_{i=1}^n (x_i-y)^2}{2\sigma_m^2}}d\x
+
\int_{\x\in\cR_{23}}
\frac1{(2\pi)^{n/2} \sigma_M^n}
\exp\Paren{-\frac{\sum_{i=1}^n (x_i-y)^2}{2\sigma_M^2}}d\x.
\]
For $n\ge2$ we transform the first integral above corresponding to the
contribution of $\cR_{21}$ from variables $\x$ to
\[
y= \frac{\sum_{i=1}^n x_i}n, \text{ and } z_j = \frac{x_j-y}{\sigma_m} \text{ for } 1\le j\le n-1,
\]
with the new variable $y$ running from $-\alpha/2$ to $\alpha/2$, and
the variables $z_1\upto z_{n-1}$ taking all possible values such that
$\sum_{j=1}^{n-1} z_i^2 +\Paren{\sum_{i=1}^{n-1} z_i}^2 \le n$.  The
Jacobian in this case is computed similar to~\eqref{eq:jac}, 
\[
\left|\frac{\partial y\z}{\partial
    \x}\right|
=
\frac{(-1)^n}{n\sigma_m^{n-1}}.  
\]
The second integral corresponding
to the contribution of the region $\cR_{23}$ is similarly transformed with variables (please note
the reuse of notation $z_j$ for later simplicity)
\[
y= \frac{\sum_{i=1}^n x_i}n, \text{ and } z_j = \frac{x_j-y}{\sigma_M} \text{ for } 1\le j\le n-1.
\]
Recalling from~\eqref{eq:inti} that 
\[
\frac{\sqrt n}{\sqrttpnmo}
\int_{z_1\upto z_{n-1}}
\!\!\!\!\!\!\!\!\!\!\!\!
\exp\Paren{ - \frac{\sum_{i=1}^{n-1}z_i^2+\Paren{\sum_{i=1}^{n-1}z_i}^2}2}
d\z
=1,
\]
we obtain that $\cR_{11}$ and $\cR_{13}$ together contribute
\[
\alpha\sqrt{\frac{n}{2\pi}}
\Paren{\frac{I_n }{\sigma_m}
+
\frac{(1-I_n)}{\sigma_M}}
\]
where for $n\ge2$
\[
I_n
\ed
\frac{\sqrt n}{\sqrttpnmo}
\int_{\substack{\sum_{i=1}^{n-1} z_i^2 +(\sum_{j=1}^{n-1} z_j)^2\le n}}
\!\!\!\!\!\!\!\!\!\!\!\!
d\z\exp\Paren{ - \frac{\sum_{i=1}^{n-1}z_i^2+\Paren{\sum_{i=1}^{n-1}z_i}^2}2}.
\]
The case $n=1$ has already been handled by Example~\ref{eg:mv}. When
$n=1$, we do not have the variables $\z$ as in the definition
above. Instead we will define $I_1=1$ for consistency with Example~\ref{eg:mv}.

The dominant contribution to the attenuation comes from $\cR_{22}$. This 
region contributes
\[
\int_{\cR_{22}}\frac{n^{n/2}e^{-n/2}}{(2\pi)^{n/2}\Paren{\sum_{i=1}^n (x_i-y)^2}^{n/2}}d\x.
\]
Note that in $\cR_{22}$, $\sum_{i=1}^n (x_i-y)^2\ge n\sigma_m^2>0$. It is also interesting
to note that this region is non-existent when $n=1$.
For $n\ge2$, we begin as in Theorem~\ref{thm:galpha}, transforming $\x$ into
\[
y= \frac{\sum_{i=1}^n x_i}n, \text{ and } z_j = {x_j-y} \text{ for } 1\le j\le n-1.
\]
We then have
\begin{align}
\nonumber
\int_{\cR_{22}}&\frac{n^{n/2}e^{-n/2}}{(2\pi)^{n/2}\Paren{\sum_{i=1}^n (x_i-y)^2}^{n/2}}d\x\\
\nonumber
&\aeq{(a)}
\int_{\substack{y,\z\\-\alpha/2\le y\le \alpha/2\\ \sigma_m^2\le \frac{\z^T\K^{-1}\z}n \le \sigma_M^2}}
\frac
{n^{n/2}e^{-n/2}}
{(2\pi)^{n/2}(\z^T\K^{-1}\z)^{n/2}} ndyd\z\\
\nonumber
&\aeq{(b)}
\int_{
\substack{
y,\ww\\
-\alpha/2\le y\le \alpha/2\\ 
\sigma_m^2\le \ww^T\ww/n \le \sigma_M^2}}
\frac
{n^{n/2}e^{-n/2}}
{(2\pi)^{n/2}(\ww^T\ww)^{n/2}} \sqrt{n} dyd\ww\\
\nonumber
&=
\alpha
\int_{
\substack{
\ww\\
\sigma_m^2\le \ww^T\ww/n \le \sigma_M^2}}
\frac
{n^{n/2}e^{-n/2}}
{(2\pi)^{n/2}(\ww^T\ww)^{n/2}}\sqrt{n} dyd\ww\\
\label{eq:two}
&\aeq{(c)}
\alpha
\frac{n^{n/2}e^{-n/2}S(n-1)}
{(2\pi)^{n/2}} 
\int_{r=\sqrt{n}\sigma_m}^{\sqrt{n}\sigma_M} \frac{\sqrt{n}}{r^2} dr\\
\nonumber
&=
\frac{\alpha\sqrt{n(n-1)}}{\pi\sqrt{2}}\Paren{\frac1{\sigma_m}-\frac1{\sigma_M}}+\cO(1).
\end{align}
Here $(a)$ follows from~\eqref{eq:zkinv} and because $|\K^{-1}|=n$
from~\eqref{eq:kinv}. To define $\ww$ in $(b)$ first note
that~\eqref{eq:zkinv} implies that $K^{-1}$ is positive definite. We
let the Cholesky decomposition of $\K^{-1}=\C^T\C$, and set
$\ww=\C\z$.  From~\eqref{eq:kinv} we have the determinant
$|C|=\sqrt{n}$ to account for the transformation of variables $z$ to $\ww$. The
equality $(c)$ follows from a transformation of $\ww$ into polar
coordinates in $n-1$ dimensions, where $S(n-1)$ is the surface area of
a sphere in $n-1$ dimensions and is equal to
\[
\frac{(n-1)\pi^{\frac{n-1}2}}{\Gamma\Paren{\frac{n-1}2+1}}.
\] 
The last line above uses the Stirling approximation which has a multiplicative approximation of $1+\cO\Paren{\frac1n}$
as specified in Lemma~\ref{lm:str}, as well as the approximation $\Paren{1+\frac1{n-1}}^{\frac{n-1}2}=\sqrt{e}+\cO\Paren{\frac1n}$. Therefore we have
the $\cO(1)$ correction term in the last line.
The Theorem follows.
\eTheorem

If we had not approximated for the Gamma function in the above proof,
we would have the precise form for attenuation. The following
Theorem~\ref{thm:exact} does exactly that---it proceeds just as
Theorem~\ref{thm:app} but leaves steps~\eqref{eq:one}
and~\eqref{eq:two} as they are without approximations. In addition,
Theorem~\ref{thm:exact} subsumes Example~\ref{eg:mv} as well.

\bTheorem
\label{thm:exact}
For all $n\ge1$,
\[
\atn(G_{\alpha,\sigma_m,\sigma_M}^n)=
\frac{\alpha n^{n/2}(n-1) e^{-n/2} }
{2^{n/2}\sqrt{\pi}\Gamma\Paren{\frac{n}2+\half}}
\Paren{\frac1{\sigma_m}-\frac1{\sigma_M}}
+
\frac{\alpha\sqrt{n}}{\sqrt{2\pi}}
\Paren{\frac{I_n}{\sigma_m}+\frac{1-I_n}{\sigma_M}}
+
\frac{n^{n/2+1} e^{-n/2} }{2^{n/2}\Gamma\Paren{\frac{n}2+1}}\ln\frac{\sigma_M}{\sigma_m}+1,
\]
where $I_1\ed1$ and 
\[
I_n
\ed
\frac{\sqrt n}{\sqrttpnmo}
\int_{\substack{\sum_{i=1}^{n-1} z_i^2 +(\sum_{j=1}^{n-1} z_j)^2\le n}}
\!\!\!\!\!\!\!\!\!\!\!\!
\exp\Paren{ - \frac{\sum_{i=1}^{n-1}z_i^2+\Paren{\sum_{i=1}^{n-1}z_i}^2}2}d\z.\eqed
\]
\eTheoremp

\section{Special cases}
\paragraph*{Constant mean, variance between $\sigma_m^2$ and $\sigma_M^2$}
Using Theorem~\ref{thm:exact}, we can also obtain the attenuation when
the mean is fixed and the variance is allowed to vary. Let
$G_{\sigma_m,\sigma_M}$ the set of all Gaussian distributions over
$\reals$ with mean 0 and whose variance lies between $\sigma_m$ and
$\sigma_M$.  As before, we denote the collection of all pdfs on
$\reals^n$ obtained by \iid sampling from a distribution in
$G_{\sigma_m\sigma_M}$ by $G_{\sigma_m\sigma_M}^n$.  \bCorollary
\label{corr:variance}
For all $n\ge1$
\[
\atn (G_{\sigma_m\sigma_M}^n) 
= 
\frac{n^{n/2+1} e^{-n/2} }{2^{n/2}\Gamma\Paren{\frac{n}2+1}}\ln\frac{\sigma_M}{\sigma_m}+1
=
\sqrt{\frac{n}\pi}\ln\frac{\sigma_M}{\sigma_m}+\cO\Paren{\frac1{\sqrt{n}}}.
\]
\Proof We obtain the above by setting $\alpha=0$ in Theorem~\ref{thm:exact}. The approximate
value comes from Theorem~\ref{thm:app}.
\eCorollary
It is easy to see that in the above result, the exact value of the mean does not matter so long
as all Gaussians have the same mean. Therefore the attenuation of the collection of all Gaussians
whose mean is $\beta$ and whose variance is in between $\sigma_m$ and $\sigma_M$ remains the same
as Corollary~\ref{corr:variance}, namely
\[
\frac{n^{n/2+1} e^{-n/2} }{2^{n/2}\Gamma\Paren{\frac{n}2+1}}\ln\frac{\sigma_M}{\sigma_m}+1,
\]
which is in turn equal to
\[
\sqrt{\frac{n}\pi}\ln\frac{\sigma_M}{\sigma_m}+\cO\Paren{\frac1{\sqrt{n}}}.
\]
\paragraph*{Constant variance $\sigma^2$, mean between $-\alpha/2$ and $\alpha/2$}
We have considered the case $\sigma^2=1$ in Theorem~\ref{thm:galpha}
already. Note that one could obtain Theorem~\ref{thm:galpha} from
Theorem~\ref{thm:exact} by setting $\sigma_m=\sigma_M=1$. For any
fixed $\sigma^2>0$ and all $n\ge1$, the attenuation of length-$n$ \iid strings from
the collection of all Gaussians with variance $\sigma^2$ and mean
between $-\alpha/2$ to $\alpha/2$ is
\[
1+\frac{\alpha}{\sigma}\sqrt{\frac n{2\pi}}
\]
by setting $\sigma_m=\sigma_M=\sigma$ in Theorem~\ref{thm:exact}.

\section{Acknowledgments}
Narayana Santhanam was supported in this research by National Science
Foundation Grants CCF-1065632 and CCF-1018984.

\section{Author contributions}
Both authors are equal contributers to this work. Both have read
the manuscript and have approved it.

\bibliographystyle{unsrt}
\bibliography{./univcod}

\end{document}